\documentstyle[a4p,12pt,amsbsy,epsf]{article}

\newcommand {\fittaubs}    {\mbox{$1.72 ^{+0.20}_{-0.19}$}}
\newcommand {\taubsys}     {\mbox{$^{+0.18}_{-0.17}$}}
 
\newcommand {\ncand}           {911}   
\newcommand {\ncanderr}        {83}   

\newcommand {\phipifitncand}   {\mbox{629}}   
\newcommand {\phipifitncanderr}{\mbox{64}}   

\newcommand {\phipifitmkkpi}   {\mbox{$1.966\pm 0.002\,\GeV$}}
\newcommand {\phipifitsigma}   {\mbox{$18\pm 2\,\MeV$}}
\newcommand {\phipifitbgfrac}  {\mbox{$0.75$}}   

\newcommand {\kskfitncand}     {\mbox{282}}   
\newcommand {\kskfitncanderr}  {\mbox{53}}   

\newcommand {\kskfitmkkpi}     {\mbox{$1.959\pm 0.004\,\GeV$}}
\newcommand {\kskfitsigma}     {\mbox{$21\pm 3\,\MeV$}}
\newcommand {\kskfitbgfrac}    {\mbox{$0.86$}}   

\newcommand {\nnphipifitncand}   {\mbox{232}}   
\newcommand {\nnphipifitncanderr}{\mbox{29}}   

\newcommand {\nnphipifitbgfrac}  {\mbox{$0.54$}}   

\newcommand {\nnkskfitncand}     {\mbox{104}}   
\newcommand {\nnkskfitncanderr}  {\mbox{17}}   

\newcommand {\nnkskfitbgfrac}    {\mbox{$0.61$}}   

\newcommand {\ncanddl}    {10633}         
\newcommand {\mlodl}      {\mbox{$50\,\MeV$}}  
\newcommand {\mhidl}      {\mbox{$200\,\MeV$}} 
\newcommand {\nGPMH}      {\mbox{3.7}}   
\newcommand {\Zzero} {\mbox{$\mathrm{ Z}^0$}}

\newcommand {\Ztocc}{\mbox{$\mathrm{ Z^0\rightarrow c\bar{c}}$}} 
\newcommand {\ztocc}{\mbox{$\mathrm{ Z^0\rightarrow c\bar{c}}$}}

\newcommand {\Ztobb}{\mbox{$\mathrm{ Z^0\rightarrow b\bar{b}}$}}

\newcommand {\zzero} {\mbox{${ z_0}$}}

\newcommand{\Rb}{\Gamma_{\rm b\bar{b}}/\Gamma_{\rm had}}
\newcommand{\Rc}{\Gamma_{\rm c\bar{c}}/\Gamma_{\rm had}}
\newcommand {\qc}          {{\mathrm c}}
\newcommand {\qb}          {{\mathrm b}}
\newcommand {\K}          {{\mathrm K}}
\newcommand {\k}          {{\mathrm K}}
\newcommand {\D}          {{\mathrm D}}
\newcommand {\B}          {{\mathrm B}}
\newcommand {\X}          {{\mathrm X}}

\newcommand {\Ds}         {\mathrm{ D_s^-}}
\newcommand {\ds}         {\mathrm{ D_s^-}}

\newcommand {\Dsm}        {\mbox{$\mathrm{ D_s^-}$}}

\newcommand {\Bu}         {{\mathrm{ B}^+}}
\newcommand {\bu}         {{\mathrm{ B}^+}}
\newcommand {\Bd}         {{\mathrm{ B^0}}}
\newcommand {\bd}         {{\mathrm{ B^0}}}
\newcommand {\Lb}         {{\Lambda_b}}
\newcommand {\Bs}         {\mathrm{B_s^0}}
\newcommand {\bs}         {\mathrm{B_s^0}}

\newcommand {\cc}         {{c\bar{c}}} 

\newcommand {\fbu}        {{f_\Bu}}
\newcommand {\fbd}        {{f_\Bd}}
\newcommand {\fcc}        {{f_{\rm c\bar{c}}}}
\newcommand {\flb}        {{f_\Lb}}

\newcommand {\KKpi}       {\rm{K}^+\mathrm{K}^-\pi^-}
\newcommand {\kkpi}       {\rm{K}^+\mathrm{K}^-\pi^-}
\newcommand {\KKp}        {\mathrm{K}^+\mathrm{K}^-\pi^-}
\newcommand {\kkp}        {\mathrm{K}^+\mathrm{K}^-\pi^-}

\newcommand {\pbs}         {p_{\mathrm{B_s}}}
\newcommand {\mbs}         {m_{\mathrm{B_s}}}

\newcommand {\pds}         {p_{\mathrm{D_s}}}
\newcommand {\pdsi}        {p_{\mathrm{D_s}}^i}
\newcommand {\mds}         {m_{\mathrm{D_s}}}
\newcommand {\lb}          {l_b}

\newcommand {\tbs}         {\tau_{\Bs}}
\newcommand {\tds}         {\tau_{\Ds}}

\newcommand{\li}           {L^i}

\newcommand{\sigli}        {\sigma_{L}^i}
\newcommand {\dEdx}       {\mbox{${\rm d}E/{\rm d}x$}}
\newcommand {\ps}         {\mbox{$\mathrm{\,ps}$}}

\newcommand {\mic}        {\mbox{$\mathrm{\,\mu m}$}}
\newcommand {\tauBs}      {\mbox{$\mathrm{\tau(B_s^0)}$}}
\newcommand {\taubs}      {\mbox{$\mathrm{\tau(B_s^0)}$}}

\newcommand {\tbg}        {\tau_{bg}}
\newcommand {\fzero}      {{f^0}}

\newcommand {\mi}         {\mbox{$m_i$}}

\newcommand {\fbgi}       {\mbox{${\cal F}^{comb}(m_i)$}}

\newcommand {\MeV}        {\mathrm{MeV}}
\newcommand {\mev}        {\mathrm{MeV}}
\newcommand {\GeV}        {\mathrm{GeV}}
\newcommand {\gev}        {\mathrm{GeV}}
\newcommand {\ts}         {\thinspace}
\newcommand {\bsym}       {\boldsymbol}

\newcommand {\Caption}[1]  {\caption[]{\small\protect{\parbox[t]{13cm}{#1} }} }
\newcommand {\downto}
        {\mbox{ \begin{picture}(14,10)
                   \put(0,10){\line(0,-1){5.0}}
                   \put(2,5){\oval(4,4)[bl]}
                   \put(1,0){\makebox(0,0)[bl]{$\rightarrow$}}
                \end{picture} }}
\begin{document}
\pagenumbering{roman}
 
\begin{titlepage}
 
  \begin{center}
    {\large\bf EUROPEAN LABORATORY FOR PARTICLE PHYSICS}
  \end{center}
 
  \begin{flushright}
    CERN-PPE/97-095\\
    July 25, 1997 \\
  \end{flushright}

 
  \vspace{1.2in}
  \begin{center}{\LARGE\bf
    A Measurement of the $\bsym\Bs$ Lifetime \\
    using Reconstructed $\bsym\Ds$ Mesons }
  \end{center}
 
  \vspace{0.3in}
  \begin{center}
    {\Large\bf The OPAL Collaboration \large}
  \end{center}
 
  \vspace{0.1in}
  \begin{abstract}
    We report a measurement  of the $\Bs$
    meson lifetime from $\Bs \to \Ds \rm{X}$ decays,  where
    $\Ds$ mesons are reconstructed in the $\Ds \to
    \mathrm{\phi\pi^-}$ and $\Ds \to\mathrm{K^{*0}K^-}$ decay channels.
    From  approximately \nGPMH~million hadronic
    $\mathrm{Z}^0$ decays recorded by the OPAL detector at LEP 
    a sample is selected containing 911 $\pm$ 83 
    candidates, of
    which $519 \pm 136$ are estimated to be from $\Bs$ meson decays.
    Fitting the distribution of the distance from the beam spot to the
    decay vertex of the $\Ds$ candidates with an
    unbinned 
    likelihood function
    we measure \vspace{0.1in} 
  
    $$   \tauBs = \fittaubs \taubsys \ps,$$
    
    where the errors are statistical and systematic, respectively.
  \end{abstract}
  \medskip

  \medskip
  \begin{center}
    (To be submitted to Zeitschrift f\"ur Physik)
  \end{center}


\end{titlepage}
\newpage
\begin{center}
{\small
K.\thinspace Ackerstaff$^{  8}$,
G.\thinspace Alexander$^{ 23}$,
J.\thinspace Allison$^{ 16}$,
N.\thinspace Altekamp$^{  5}$,
K.J.\thinspace Anderson$^{  9}$,
S.\thinspace Anderson$^{ 12}$,
S.\thinspace Arcelli$^{  2}$,
S.\thinspace Asai$^{ 24}$,
D.\thinspace Axen$^{ 29}$,
G.\thinspace Azuelos$^{ 18,  a}$
% ...
}
\end{center}
{\small
% ...
$^{  2}$Dipartimento di Fisica dell' Universit\`a di Bologna and INFN,
I-40126 Bologna, Italy
\newline
% ...
$^{  5}$Cavendish Laboratory, Cambridge CB3 0HE, UK
\newline
% ...
$^{  8}$CERN, European Organisation for Particle Physics,
CH-1211 Geneva 23, Switzerland
\newline
$^{  9}$Enrico Fermi Institute and Department of Physics,
University of Chicago, Chicago IL 60637, USA
\newline
% ...
$^{ 12}$Indiana University, Department of Physics,
Swain Hall West 117, Bloomington IN 47405, USA
\newline
% ...
$^{ 16}$Department of Physics, Schuster Laboratory, The University,
Manchester M13 9PL, UK
\newline
% ...
$^{ 18}$Laboratoire de Physique Nucl\'eaire, Universit\'e de Montr\'eal,
Montr\'eal, Quebec H3C 3J7, Canada
\newline
% ...
$^{ 23}$Department of Physics and Astronomy, Tel Aviv University,
Tel Aviv 69978, Israel
\newline
$^{ 24}$International Centre for Elementary Particle Physics and
Department of Physics, University of Tokyo, Tokyo 113, and
Kobe University, Kobe 657, Japan
\newline
% ...
$^{ 29}$University of British Columbia, Department of Physics,
Vancouver BC V6T 1Z1, Canada
\newline
% ...
$^{  a}$ and at TRIUMF, Vancouver, Canada V6T 2A3
\newline
% ...
\newline
}
\newpage
\pagenumbering{arabic}
\section{Introduction}
\label{sec:intro}

The lifetimes of b-flavored hadrons are related to both the strength
of the b quark coupling to c and u quarks, and to the dynamics of b
hadron decay.  The spectator model assumes that the light quarks in b
and c hadrons do not affect the decay of the heavy quark, and thus
predicts the lifetimes of all b-hadrons to be equal.  For charm
hadrons this prediction is inaccurate; the measured $\mathrm{D^+}$
lifetime is approximately 2.5 times that of the $\mathrm{D^0}$ and
more than twice that of the $\Ds$~\cite{PDG}.  More sophisticated
models predict that the differences among b-hadron lifetimes should be
much smaller than those in the charm system, because of the larger
mass of the b quark~\cite{HQET,BIGI,NEUBERT}.  Bigi et al.~\cite{BIGI}
predict a difference in lifetime between the $\B^+$ and $\Bd$ meson
of several percent, and between the $\Bs$ and $\Bd$ mesons of about
$1\%$.  Although some assumptions in reference~\cite{BIGI} have been
questioned by Neubert and Sachrajda~\cite{NEUBERT}, there is agreement
that the models predict a 
difference between the $\Bs$ and $\Bd$ lifetimes 
of the order of $\pm1\%$.

The first measurements of the $\Bs$
lifetime~\cite{bslifetime,opalbslife} were made using correlated
$\Ds$-lepton pairs that primarily result from the semileptonic decay
of the $\Bs$. However, the small $\Bs$ semileptonic branching ratio limits
the statistical power of this channel.  More inclusive techniques can
be used to circumvent this limitation.
 
We present a new measurement of the lifetime of
the $\Bs$ meson in which only the $\Ds$ mesons are reconstructed.
The two decay channels used are:\footnote{Charge
  conjugate modes are always implied.  Also, unless otherwise noted, K
  and $\pi$ always refer to charged particles.} \vspace{-20pt}
\begin{tabbing}
  \hspace{3cm} \= \hspace{4.7cm} \= \hspace{2cm} \= \hspace{3cm} \\
  \> $\Bs \to \Dsm\, \mathrm{X}$ \>
  \> $\Bs \to \Dsm\, \mathrm{X}$ \\
  \> $\phantom{\Bs \to \hspace{5pt}} \downto \K^{*0}\, \K^-$ \> and
  \> $\phantom{\Bs \to \hspace{5pt}} \downto \phi\, \pi^-$ \\
  \> $\phantom{\Bs \to \downto} \downto \ \K^+\,\pi^-$ \>
  \> $\phantom{\Bs \to \downto} \downto \ \K^+\,\K^-$.
\vspace{-5pt}
\end{tabbing}
The $\Bs$ meson is not the only source of $\Ds$ mesons.
Significant numbers of $\Ds$ mesons are  produced in the decays 
of other b-hadrons as well as in $\Ztocc$ events.
The level and shape of this background is evaluated using Monte Carlo
data and measured branching fractions. The background from random
tracks combination is evaluated using the side band technique.

To measure the $\Bs$ lifetime, we reconstruct the decay
vertex of the $\Ds$ and determine the distance from the beam spot to
this point.  For $\Ds$ mesons that result from the decay of b-hadrons, 
this ``decay length'' has contributions both from the b-hadron and 
$\Ds$ decay lengths. The bias arising from the angle between the $\Bs$
meson and the $\Ds$ meson direction is very small and is not affecting the
result of this analysis. 
An unbinned maximum likelihood fit is performed using
the reconstructed decay lengths, their errors and the $\Ds$ momentum to
extract the mean $\Bs$ lifetime.
The following sections describe the OPAL detector, the selection of
$\Ds$ candidates, the vertex topology of the events, the determination
of the $\Bs$ decay length, the estimation of the $\Bs$ energy, the
lifetime fit, the results, and the systematic errors.
 
\section{The OPAL Detector}
\label{sec:detector}
 
The OPAL detector is described in
reference~\cite{opaldet-opalsi-opalsi2}. The central tracking system
is composed of a silicon microvertex detector, a precision vertex drift
chamber and a large volume jet chamber surrounded by a set of chambers to
measure the $z$-coordinate of tracks ($z$-chambers)\footnote{The
  coordinate system is
  defined such that the $z$-axis follows the electron beam direction
  and the $x$-$y$~plane is perpendicular to it with the $x$-axis lying
horizontally.  The polar angle~$\theta$~is defined
  relative to the $+z$-axis, and the azimuthal angle~$\phi$~is defined
  relative to the $+x$-axis.}.  These detectors are
located inside a solenoid. The detectors outside the solenoid
consist of a time-of-flight scintillator array and a lead glass
electromagnetic calorimeter with a presampler, followed by a hadron
calorimeter consisting of the instrumented return yoke of the magnet,
and several layers of muon chambers.  Charged particle types are identified
by their specific energy loss, $\dEdx$, in the jet chamber.  Further
information on the performance of the tracking and $\dEdx$ measurements
can be found in reference~\cite{opaljet-hid}.
  
\section{$\bf \Ds$ Candidate Selection}
\label{sec:selection}
 
This analysis uses data collected during the 1991--1995 LEP running
periods at center-of-mass energies within $\pm3\,\GeV$ of the
$\mathrm{Z}^0$ mass.  After the standard hadronic event
selection~\cite{opalmh} and detector performance requirements, a
sample of \nGPMH~million events is selected.
Charged tracks and electromagnetic clusters not associated with a
charged track are grouped into jets using the JADE E0 recombination
scheme~\cite{bib-JADE} with a  $y_{\rm cut}$ value of 0.04.  Tracks from
identified secondary vertices, $\Lambda$ and
$\mathrm{K^0_S}$ decays or $\gamma$ conversions, are excluded from the $\Bs$
candidate selection.

Simulated event samples were generated using the JETSET 7.4 Monte
Carlo program~\cite{jetset}, together with a program to simulate the
response of the OPAL detector~\cite{opalmc}. The Monte Carlo sample
 includes approximately 4 million simulated multihadronic $\Zzero$
decays and one million $\Ztobb$ decays (the equivalent of about 4.5
million multihadronic decays). In addition, three special 
Monte Carlo samples were generated in which each event
contains at least one $\Ds$ decaying in the channels of
interest.
The parameter optimisation used in this simulation is described 
in~\cite{jetset}.
For each of the following decay channels, 20000 events were
generated: $\Ds\to\Ds X$, $\Ds$ from b-hadrons other than $\Ds$
decays and $\Ds$ from $\Ztocc$.

\subsection{$\bf \Dsm \to \KKpi$ selection}
 
The  $\Ds$ meson is  reconstructed in the decay chains
$\Dsm\to\K^{*0}\K^-$ in which the $\K^{*0}$ decays into a
$\K^+\pi^-$, and $\Dsm\to\phi\pi^-$ where the $\phi$
subsequently decays into $\K^+\K^-$.  

Tracks forming the $\Ds$ candidates are required to be in the same jet
and to have the appropriate charge combination. At least two of the three
candidate tracks are required to have good $\theta$ measurements
either from the $z$-chambers or from a measurement of the track
endpoint as it exits the main jet chamber.  Similarly, to reject
poorly reconstructed candidates, at least two of the three tracks are
required to have hits in the silicon microvertex detector.

To reduce the combinatorial background, the tracks forming the $\Ds$
are subject to particle identification requirements.  For candidate pion
tracks, the probability for the measured $\dEdx$ value to be
consistent with the pion hypothesis is required to be greater than
1\%.  For candidate kaon tracks, if the observed energy loss of a kaon
candidate is less than the mean $\dEdx$ expected for a kaon, the
probability of consistency with the kaon hypothesis is required to be 
greater than
1\%, and greater than 3\% otherwise.  If both kaon candidates have
energy losses greater than the mean $\dEdx$ expected for a kaon, the
product of the two $\dEdx$ probabilities is required to be greater
than 0.02.  These tighter requirements reduce the background from pion
tracks, for which the mean $\dEdx$ value is above that of kaons.
Furthermore, one of the kaon candidates must satisfy a pion-rejection
criterion by having an observed  $\dEdx$ less than that expected
for a pion, and a probability of less than 10\% that the $\dEdx$ is
consistent with a pion hypothesis.

For the $\K^{*0}\K^-$ mode, the invariant mass of the $\K\pi$
combination is required to satisfy $0.865 <m_{\K\pi} < 0.925\,\GeV$.
In the $\phi\pi$ mode, the width of the $\k^+\k^-$ peak is dominated by
detector resolution and the $\K^+\K^-$ invariant mass is required to 
satisfy $1.010 < m_{\K\K} < 1.030\,\GeV$.  
The momenta of the kaons are required to be greater
than $2\,\GeV$ and that of the pion must exceed $1\,\GeV$.

To reduce further the $\D^-\to\k^{*0}\pi^-$ background in the
$\k^{*0}\k^-$ mode, the kaon candidate originating directly from the
$\Ds$ decay is subject to a tighter, 5\%, $\dEdx$ requirement.  Also,
in this mode both kaons must meet the pion-rejection criterion
described above.

We require that the $\Ds$ momentum divided by the beam energy, $x_{\Ds}$,
is greater than 0.20 to reduce random track combinations, and
less than 0.60 to reduce the $\Ds$ contribution from $\Ztocc$ events.
 
The differences between the angular distributions of $\Ds$ decays and
those of combinatorial background are exploited to enhance the signal
purity as follow.
The $\Ds$ is a spin-0 meson and the final
states of both decay modes consist of a spin-1 ($\phi$ or $\K^{*0}$)
meson and a spin-0 ($\pi^-$ or $\K^-$) meson.  The $\ds$ signal is
expected to have no dependence on $\cos\theta_p$, where ${\theta_{p}}$
is the angle in the rest frame of the $\ds$ between the spin-0 meson
direction and the $\ds$ direction in the lab frame.  However, the
$\cos\theta_p$ distribution of random combinations peaks in the
forward and backward directions.  It is therefore required that
$|\cos\theta_p| < 0.8$ for both modes.  The distribution of
$\cos \theta_v$, the angle in the rest frame of the spin-1 meson between
the direction of the final state kaon from the decay of the spin-1
meson and the $\ds$ direction, is proportional to $\cos^2\theta_v$ for
$\ds$ decays.  The $\cos\theta_v$ distribution of the random $\kkp$
combinations in the data is, however, approximately uniform.  Therefore
it is required that $|\cos\theta_v| > 0.6$ (0.4) for the $\K^*\K$
($\phi\pi$) mode.
 
\subsection{Decay length determination}
 
The $\Ds$ decay vertex is reconstructed in the $x$-$y$ plane by
fitting the $\Ds$ candidate tracks to a vertex.  To reject very poorly
reconstructed vertices, the $\chi^2$ of the vertex fit is required to
be less than 10 (for 1 degree of freedom).

The beam spot position is measured using charged tracks in the opal data with a
technique that follows any significant shifts in the
position during a LEP fill~\cite{taulife}.  The intrinsic width of the
beam spot in the $y$ direction is about $8\mic$.  the width in the $x$
direction is directly measured using $\mu^+ \mu^-$ events and varies
between $100\mic$ and $160\mic$, depending on the LEP optics.
 
The distance from the $\Ds$ decay vertex to the beam spot is
determined in the $x$-$y$ plane. This
distance is converted into three dimensions using the polar angle
of the $\KKp$ momentum vector.  
Typical decay length errors are about $300\mic$, with only a small
contribution coming from uncertainties in the position of the
interaction point within
the beam spot.  Rejecting candidates with decay length errors
greater than 3~mm helps to reduce the effects of 
poorly measured tracks.


\subsection{$\bf \Ds$ selection results}
 
\begin{figure}[btp]
  \begin{center}
  \begin{minipage}{0.9\textwidth}
     \begin{center}
       \epsfxsize=\textwidth
       \epsffile{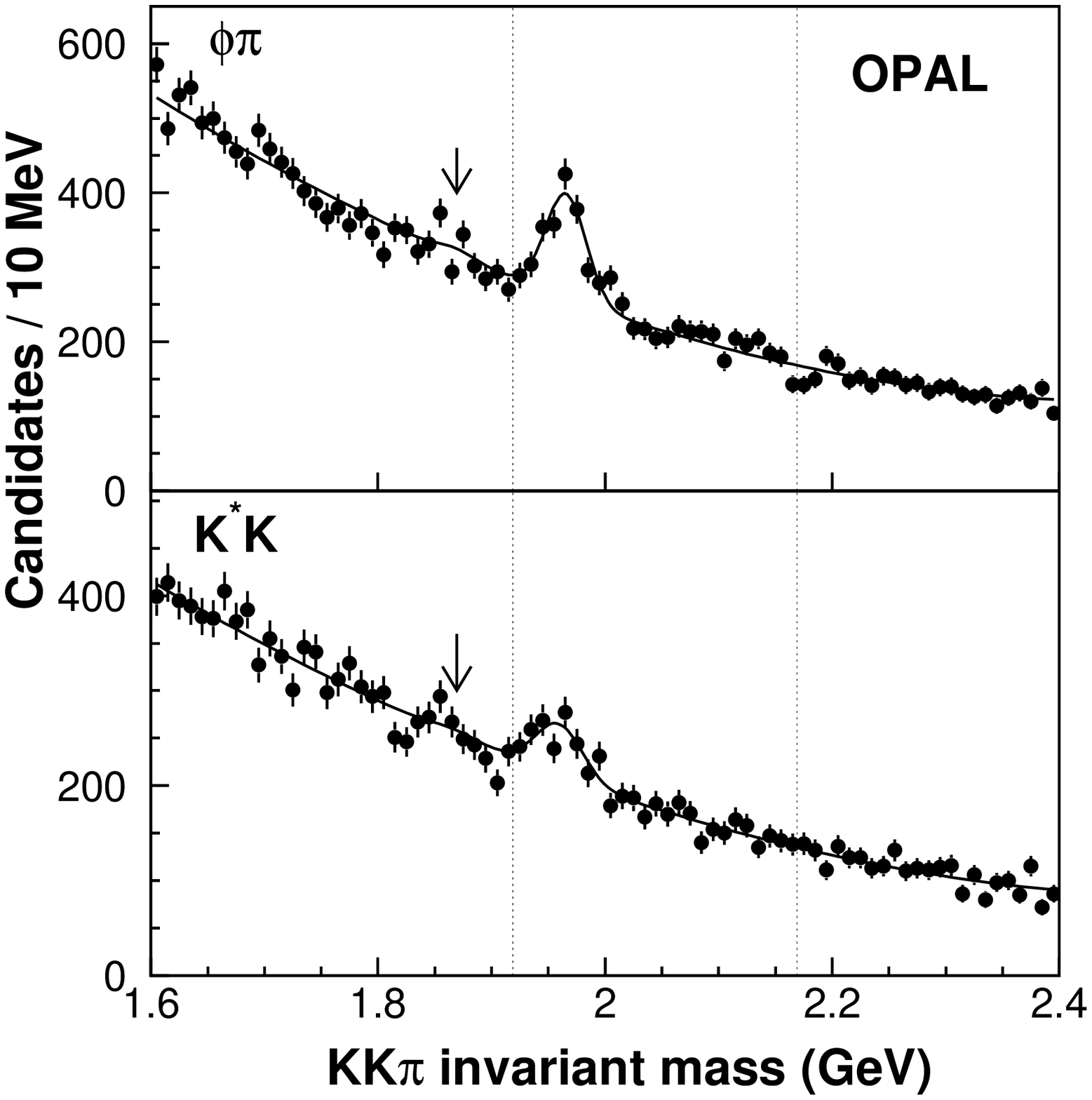}
     \end{center}
     \vspace{-0.6in}
  \end{minipage}
  \begin{minipage}{0.9\textwidth}
  \Caption{Results of the standard selection.
      Top: $\kkp$ invariant mass distribution
      for $\phi\pi$ combinations along with the fitted distribution.
      Bottom: $\kkp$ invariant mass distribution for $\k^*\k$
      candidates, along with the fitted distribution.
      The dotted lines indicate the region within which 
      candidates are used in the lifetime fit. 
      The arrow indicates the $\rm D^{-}$ peak position.}
     \label{fig:dsmass}
  \end{minipage}
  \end{center}
\end{figure}

The $\KKp$ invariant mass distribution for all candidates that 
pass the selection is
shown in figure~\ref{fig:dsmass}.
A likelihood fit to the measured $\kkp$ invariant mass of the $\ds$
candidates is performed separately for the two decay channels.  The
$\kkp$ mass distribution is parameterized as the sum of a quadratic
term to account for random combinatorial background (which also is
observed to describe accurately the background in the simulated data
samples), a Gaussian function which describes the mass peak of the
reconstructed $\ds$ signal, and another Gaussian to account for a
$\D^-\to\KKpi$ contribution.  This last term has a mean fixed to the
$\D^-$ mass of $1869.3\,\mev$~\cite{PDG} and the width constrained
to be the same as that of the $\ds$ signal. This second Gaussian is
included to avoid biasing the estimate of the combinatorial
background.  The results of this fit to the $\kkpi$ invariant mass
spectra are shown in table~\ref{tab-mfit}.  A total of $\ncand
\pm \ncanderr$ $\ds$ candidates are found.
\begin{table}[thb]
  \centering
  \begin{tabular}{|l|c|c|}
    \hline
      source  &  $\phi\pi$ & $\k^*\k$ \\
    \hline
      number of $\ds$ candidates & 
        $\phipifitncand \pm \phipifitncanderr$  &  
        $\kskfitncand   \pm \kskfitncanderr  $  \\
    \hline
      fitted $\ds$ mass & 
        \phipifitmkkpi  & \kskfitmkkpi \\
    \hline
      fitted $\ds$ sigma & 
        \phipifitsigma  & \kskfitsigma \\
    \hline
      background fraction &
        \phipifitbgfrac & \kskfitbgfrac \\
    \hline
      total $\chi^2$ over 80 bins &
       90.0            & 80.0 \\
    \hline
  \end{tabular}
  \Caption{$\kkpi$ invariant mass spectra fitting results. The fitted
mass is in agreement with the value of $1.9685 \pm 0.0006 {\rm MeV}$ 
from~\cite{PDG}. The ``background fraction'' is defined as the fitted 
number of combinatorial background candidates divided by the total 
number of candidates within 2$\sigma$ of the fitted $\Ds$ mass.} 
  \label{tab-mfit}
\end{table}

\subsection{Composition of the $\bsym\Dsm$ signal}
\label{sec:ds-comp}

The $\bs$ meson is not the only source of $\ds$ mesons.  Significant
numbers of $\ds$ mesons are produced in other b-hadron
decays as well as in $\Ztocc$ events, collectively referred to as
`physics background'.  In what follows, the contributions of these
three components to the observed $\ds$ spectra are estimated.

The fractions of $\ds$ mesons produced from each of these sources are
extracted using the results in reference~\cite{opalcprod}.  The
measured $\ds$ production rates separated into flavour 
are $f_{\rm b} =
\Rb\cdot f(\qb\to\ds)\cdot Br(\Ds \to \phi \pi^-) = 
(0.166\pm0.018\pm0.016)\%$ and $f_{\rm c} =
\Rc\cdot f(\qc\to\ds)\cdot Br(\Ds \to \phi \pi^-) = 
(0.056\pm0.015\pm0.007)\%$.  Thus,
$f_{\rm b}/(f_{\rm c}+f_{\rm b})=(75\pm9)\%$ of produced $\ds$ mesons
are  from $\Ztobb$ events
and the remaining $(25\pm9)\%$ are from $\ztocc$
events.



The fraction of $\ds$ mesons from b-hadrons in our signal is estimated
using the following production rates: $f(\qb\to\Bs) = 0.112^{+0.018}_{-0.019}$ 
\cite{PDG} and 
$f(\qb\to\B) =0.378\pm0.022$
\cite{PDG}  (where `B' is either $\bu$ or $\bd$).
Assuming that b-baryons decay to $\ds$ mesons with the same branching
fractions as the non-strange b-mesons ($0.086\pm0.016$ \cite{PDG}) and
using the inclusive branching ratio of $\bs$ to $\ds$, measured to be
$0.87\pm0.31$ \cite{PDG}, we estimate that $(56\pm11)\%$ of $\ds$ mesons
from b-hadron decay are from $\bs$. Thus, $(42\pm10)\%$ of the $\ds$ 
mesons come from
$\bs$ decays, $(33\pm9)\%$ come from other b-hadron decays and $(25\pm9)\%$
from $\ztocc$ events.  Assuming the production rates of $\bu$ and
$\bd$ are the same and that they have equal branching ratios to $\ds$,
each of these non-strange b-mesons accounts for $(14\pm4)\%$ of the
$\ds$ production with the remaining $(5\pm2)\%$ of $\ds$ produced from
b-baryons.

The special  Monte Carlo samples described above were used to
determine the contribution of each of these channels i the
reconstructed sample.  The ratio of the
efficiency to reconstruct a $\ds$ meson from b-hadron decay other than
$\bs$, divided by the efficiency to reconstruct a $\ds$ meson from a $\bs$
decay is $0.82\pm0.02$, where the error is due to the limited
statistics of the simulated data samples.  The principal reason that
this ratio is less than unity is that the momentum spectrum of 
$\ds$ mesons from the decay of b-hadrons other than $\bs$ is softer 
than that of $\ds$ mesons from $\bs$ decay.  The ratio of
the efficiency to reconstruct a $\ds$ meson from a $\ztocc$ event
divided by the efficiency to reconstruct a $\ds$ from a $\bs$ decay is
$0.67\pm0.01$.  The upper cut on the scaled energy of the $\ds$
($x_{\ds} < 0.6$) preferentially rejects $\ds$ from $\ztocc$, 
which tend to have 
higher momentum.  the sources of $\ds$ production are
summarized in table~\ref{tab-physbg}.
\begin{table}[thb]
  \centering
  \begin{tabular}{|l|c|c|}
    \hline
      source  &  fraction of produced $\ds$ & fraction of
      reconstructed $\ds$ \\
    \hline
      $\ztocc$        & $25\pm9$ & $17\pm6$ \\
    \hline
      $\bu$           & $14\pm4$ & $11\pm3$ \\
    \hline
      $\bd$           & $14\pm4$ & $11\pm3$ \\
    \hline
      b-baryons       &  $5\pm2$ &  $4\pm2$ \\
    \hline
      $\bs$           & $42\pm10$ & $57\pm14$ \\
    \hline
  \end{tabular}
  \Caption{Estimated $\ds$ signal composition. 
    The errors include those from the
    measured branching ratios and the statistical uncertainty
    from the Monte Carlo samples used to estimate the relative
    reconstruction efficiencies.}
  \label{tab-physbg}
\end{table}

Monte Carlo events were used to study the background from events where
the three $\kkp$ candidate tracks come either from the same fully
reconstructed charm hadron for which a pion or a proton has been
mis-identified as a kaon, or from a partially reconstructed charm
hadron.  The resulting $\kkp$ invariant mass distribution in the
region around the $\ds$
mass is similar to that of the combinatorial background. Such events
are therefore, considered to contribute to the combinatorial background.

Since $(57\pm14)\%$ of the reconstructed $\ds$ meson decays are
calculated to be from  $\bs$ decays, $519\pm 136$ $\ds$ candidates
are attributed to $\bs$ decay from the $\ncand \pm \ncanderr$
candidates resulting from the fit to the $\kkp$ invariant spectra.
 
\section{The $\bsym \bs$ Lifetime Fit}
 
To extract the $\bs$ lifetime from the measured decay
lengths, an unbinned maximum likelihood fit is performed using a
function that accounts for both the $\ds$ signal and the background
components of the sample. In the part of the likelihood
function describing events from b-hadron decays, 
the observed decay lengths depend on the $\bs$ lifetime.
 
The form of the likelihood function for the candidates in the $\ds$
signal from $\bs$ decays is described in terms of the
probability for observing a combined $\ds$ and $\bs$ decay length,
$\li$, given a measurement error $\sigli$, the momenta of the $\ds$
and $\bs$, and the mean lifetimes of these mesons.




The likelihood function has components which describe the different
sources of $\ds$ mesons in the signal and in the
combinatorial background.  This follows closely the method used in
previous opal $\bs$ lifetime analyses \cite{opalbslife}.

The likelihood function which accounts for  $\ds$ mesons from $\bs$ decays
is constructed as a convolution of two exponential functions to describe the
$\ds$ and $\bs$ decay lengths, convoluted with a function to describe
the probability of having a particular $\bs$ momentum ($\pbs$) and
Gaussian functions to describe the measured decay length resolution.
This can be expressed as:
\begin{center}
${\cal L}_i^{\bs}(\li \mid \tbs ,\sigli, \pdsi, s_1,s_2,f_2) 
= \hspace{7cm}$
\end{center}
\begin{equation}
 \int_0^\infty {\mathrm d}l 
   \int_0^{\pbs(max)}{\mathrm d}\pbs  
    {\cal R}(\li \mid l,\sigli,s_1,s_2,f_2) \,
    {\cal B}(\pbs \mid \pdsi) \,
    {\cal P}(l \mid \tbs, \tds, \pbs, \pdsi) \, 
\end{equation}
where $\pbs(max) = 45 \gev$ and ${\cal R}$ is given by
\begin{equation}
  {\cal R}(\li \mid l,\sigli,s_1,s_2,f_2) =
   (1-f_2) \, {\cal G}(\li \mid l, \sigli, s_1) + 
   f_2 \, {\cal G}(\li \mid l, \sigli, s_2)\ .
\end{equation}
The function ${\cal g}$ is a Gaussian function which describes the 
probability to
observe a decay length, $\li$, given a true decay length $l$ and the
measurement uncertainty $\sigli$ and scale factors on this error,
$s_1$ and $s_2$.  Two scale factors are employed to describe both the
majority of tracks for which the measured decay length uncertainty is
a good estimate and the small fraction, $f_2$, of mis-measured candidates in
which the measured decay length uncertainty is an underestimate.  The
scale factors $s_1$ and $s_2$, as well as the fraction of mis-measured
candidates, $f_2$, are free parameters in the lifetime fit.  ${\cal
  B}$ is the probability of a particular 
$\bs$ momentum for an
observed $\ds$ momentum,$\pds$.  This probability is determined from Monte
Carlo events by forming distributions of the ratio $\pbs/\pds$.  Six
such distributions are formed, depending on the value of $\pds$, since at
higher values of $\pds$ the range of potential values of $\pbs$ is
more tightly constrained than for lower $\pds$ candidates.  ${\cal P}$
is the probability for the $\ds$ to decay at a distance $l$ from the
$e^+e^-$ interaction point, given $\bs$ and $\ds$ lifetimes $\tbs$ and
$\tds$ and momenta $\pbs$ and $\pdsi$.  This function is constructed as:
\begin{equation}
    {\cal P}(l \mid \tbs, \pbs, \pds) = 
    \frac{\mds}{\tds\pds} 
    \frac{\mbs}{\tbs\pbs} 
      \int_0^l 
        \exp \left[\frac{-l'\cdot \mbs}{\tbs\pbs}\right]
        \exp \left[\frac{-(l - l')\cdot\mds}{\tds\pds}\right] {\mathrm d}l'
\end{equation}
where $\tbs\pbs/\mbs$ and $\tds\pds/\mds$ are the mean
decay lengths for the given momenta ($\pbs$ and $\pds$) mean lifetimes
($\tbs$ and $\tds$) and masses ($\mbs$ and $\mds$). 

Similar functions are employed for the other $\ds$ meson sources.  For
$\ds$ mesons from other b-hadron decays ($\bu$, $\bd$ and b-baryons),
the world average lifetime for each species of b-hadron \cite{PDG} is
used in the likelihood function and a slightly different boost
function, ${\cal B}$, is employed.  For $\ds$ mesons produced in
$\ztocc$ events, the function ${\cal P}$ is a single exponential
function with the decay constant $\tds\pds/\mds$.  The total
likelihood function containing the contributions for all sources of
$\ds$ mesons, is formed by combining the likelihood functions for each
of the sources of $\ds$ mesons with the fixed fractions listed in
table~\ref{tab-physbg}.\footnote{The $\ds$ momentum dependence on
  The fractions $\fcc$, $\fbu$, $\fbd$ and $\flb$ has been neglected.
  This omission was studied in simulated data and found not to produce
  a noticeable bias in the resulting lifetime.} This likelihood is
written as:
\begin{eqnarray}
 {\cal L}_i^{\ds}(\li \mid \tbs ,\sigli, \pdsi) & = &
  ( 1 - \fcc - \fbu - \fbd - \flb ) {\cal L}_i^{\bs} +
\nonumber \\ 
& & \fcc \cdot {\cal L}_i^{\cc} +
  \fbu \cdot {\cal L}_i^{\bu} +
  \fbd \cdot {\cal L}_i^{\bd} +
  \flb \cdot {\cal L}_i^{\lb}  \ . 
\end{eqnarray}

The functional form of the likelihood function, ${\cal L}^{\rm comb}$,
used empirically to parameterize the combinatorial background, is
composed of an exponential convolved with the $\bs$ boost function,
${\cal B}$, a fraction with no lifetime and the same double-Gaussian
resolution function as the signal.  This is expressed as:
\begin{center}
${\cal L}_i^{\rm comb}(\li \mid \tbg,\fzero,\sigli,\pdsi,s_1,s_2,f_2)
=\hspace{6cm}$
\end{center} 
\begin{equation}
\int_0^\infty {\mathrm d}l 
   \int_0^{\pbs({\rm max})}{\mathrm d}\pbs  
    {\cal R}(\li\mid l,\sigli,s_1,s_2,f_2)
    {\cal B}(\pbs \mid \pdsi) 
    {\cal P}_{bg}(L \mid \tbg, \fzero, \pbs)  
     \ ,
\end{equation}
where the parameters describing the resolution, $s_1$, $s_2$ and $f_2$,
are the same as used in the likelihood terms that describe the $\ds$
signal and
\begin{equation}
   {\cal P}_{bg}(l \mid \tbg, \fzero, \pbs) =
 (1-\fzero)\frac{\mbs}{\tbg\pbs}\exp\left[\frac{-l \cdot \mbs}{\tbg\pbs}\right]
           + \fzero\delta(l)\  \ .
\end{equation}
The fraction of background with no lifetime, $\fzero$, as well as the
characteristic lifetime of the background, $\tbg$, are free parameters
in the fit.  

The combinatorial background in the event sample is taken into account
by fitting for it simultaneously with the $\ds$ signal.  
The probability that a candidate has arisen from a combination of
background tracks, $\fbgi$, is determined as a function of the
$\kkp$ invariant mass of each candidate, $\mi$, using the results of
the fit to the $\kkp$ invariant mass spectrum.

Thus, the full likelihood for candidate $i$ is:
\begin{equation}
 {\cal L}_i(\li \mid \tbs ,\sigli, \pdsi, \mi) \ = \
  [ 1 - \fbgi]\cdot{\cal L}_i^{\ds} +
  \fbgi\cdot{\cal L}_i^{comb} \ .
\end{equation}
In total six parameters are free in the fit: the $\bs$ lifetime, the
parameters describing possible scale factors on the decay length
error ($s_1$, $s_2$ and $f_2$) and the parameters describing the
combinatorial background ($\fzero$ and $\tbg$).

The lifetime fit uses the $\ncanddl$ events found in the region from
$\mlodl$ below the fitted $\ds$ mass to $\mhidl$ above it (see
figure~\ref{fig:dsmass}).  From studies on simulated data, it is found
that the lifetime properties of the combinatorial background in this
region accurately reflect the combinatorial background around the
$\ds$ mass.  Furthermore, this avoids the region below the $\ds$ mass
in which the number of $\kkp$ candidates from $\mathrm{D^-}$ decays
and contributions from other D meson decays (e.g. reflections and
partially reconstructed decays) are potentially significant.  This fit
finds $\taubs = \fittaubs \ps$, where the error is statistical only,
the values of the other free parameters of the fit are in table~\ref{tab-fit}.
\begin{figure}[p]
  \begin{center}
  \begin{minipage}{0.9\textwidth}
    \begin{center} 
      \epsfxsize=0.95\textwidth
      \epsffile{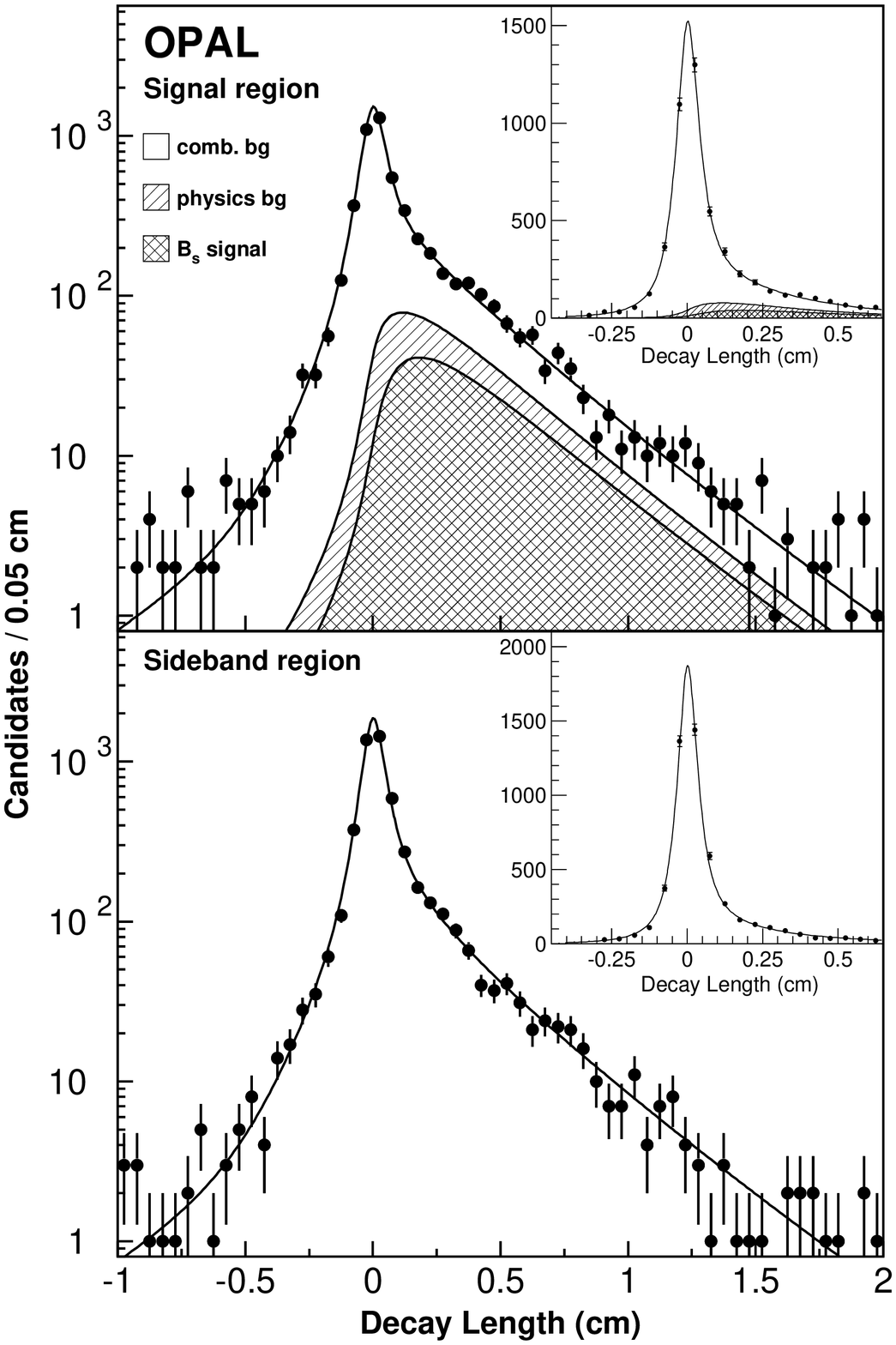}
     \end{center} 
    \vspace{-0.6in}
  \end{minipage}
  \begin{minipage}{0.9\textwidth}
    \vspace{0.in} \Caption{Top: decay length distribution within a
      $\ds$ mass region of $\pm 50\,\mev$ around the fitted $\ds$
      mass (signal region).  The single-hatched area represents the
      contribution from physics background, the unhatched area the
      combinatorial background and the cross-hatched area is the
      contribution from $\bs$ decays.  Bottom: decay length
      distribution for events outside the $\ds$ mass region (sideband
      region), namely from 50 to 200~MeV above the $\ds$ mass. }
     \label{fig:datafit}
  \end{minipage}
  \end{center}
\end{figure}
\begin{table}[thb]
  \centering
  \begin{tabular}{|l|c|}
    \hline
      parameter  &  fit results \\
    \hline
     $f_2$            & $0.07\pm0.01$ \\
    \hline
      $s_1$           & $0.88\pm0.01$ \\
    \hline
      $s_2$           & $4.69\pm0.26$ \\
    \hline
      $f^0$           & $0.77\pm0.01$ \\
    \hline
     $\tbg$           & $1.61\pm0.07$ \\
    \hline
  \end{tabular}
  \Caption{Final values of the free parameters in the fit.}
  \label{tab-fit}
\end{table}
The fitted values of the parameter are describing the decay length
resolution are consistent with the understanding of the OPAL tracking 
performances.
The decay length distributions are shown in figure~\ref{fig:datafit}
separately for candidates with $\kkp$ invariant mass within
$50\,\mev$ of the fitted $\ds$ mass (the ``signal region''), and for
those candidates outside this mass window (the ``sideband region'').
These illustrate the quality of the fit in regions dominated by $\ds$ signal
events and by combinatorial background, respectively.  The curves in
figure~\ref{fig:datafit} represent the sum of the decay length
probability distributions for each event.  Using the 42 bins that are
expected to contain at least five candidates (as predicted by the
lifetime fit), a total $\chi^2$ of 54.7 is found for the sum of the
signal and sideband decay length distributions.  For
the positive decay length bins, a total $\chi^2$ of 24.0 for
29 bins is observed.  These $\chi^2$ values and plots shown in
figure~\ref{fig:datafit} indicate that the fitted functional forms
provide a good description of the data for both signal and background.
It should be stressed that the fit is to the unbinned data.
 
\section{\label{sec-check} Checks of the Method}

A number of different checks have been made to investigate potential
biases in the method of selecting and fitting the signal.

\subsection{Potential bias in the selection and fitting procedure}

Tests were performed on several samples of simulated data to check for
biases in the selection and fitting procedures.  The first tests
involved a toy Monte Carlo program which generated decay length data
for the signal $\ds$ decays and combinatorial background.  For each
$\ds$ signal candidate from a $\bs$ decay, this simulation generated
$\bs$ and $\ds$ decay times from exponential distributions with the
means set to known values.  The $\bs$ and $\ds$ momenta were chosen
from a spectrum based on the full Monte Carlo simulation.  The $\bs$
and $\ds$ decay lengths were then calculated and combined to give the
true candidate decay length, which was then smeared by a resolution
function. physics and combinatorial backgrounds were 
generated through a similar procedure.  Many fits
were conducted over wide ranges of $\bs$ lifetimes with different
levels and parameterizations of the backgrounds.  The result of these
studies shows that any bias in the fitted $\bs$ lifetime is 
less than 0.5\% and that the statistical precision of the fit to 
data is consistent with the sample size and
composition.

To verify that the $\ds$ selection does not bias the
reconstructed sample, a lifetime measurement was made from 20000 
$\bs\to \ds x$ Monte Carlo decays into the two channels of this
analysis, using a $\bs$ lifetime of 
$1.60\ps$.  The mean lifetime of the selected
$\ds$ sample was  $1.64\pm0.04\ps$, consistent with
the expectation that there is no bias from the selection procedure.
The lifetime obtained by fitting this same sample
was $1.65\pm0.05\ps$.

To investigate the effects of the combinatorial background on the
lifetime fit, the same selection and fitting procedure has been
applied on a Monte Carlo sample of 4 million multihadronic $\zzero$
decays.  Due to the choice of branching ratios used to produce this
sample of simulated events, there are fewer reconstructed $\ds$ signal
candidates than we observe in the data. The fitted lifetime has
been found to be $1.73\pm0.29\ps$, which is consistent with the
generated $\bs$ lifetime of $1.6\ps$ within the statistical power of
this sample.  If the signal events from
the 20000 $\bs \to \ds x$ decays described above are added to the
simulated data, the resulting 
sample is of similar purity to that found by the tight neural net
selection.  The fitted lifetime of this pure  sample is
$1.65\pm0.07\ps$,  again in good agreement with the 
true lifetime of the sample.

The lifetime fit has also been repeated as for the standard result,
except that the $\phi\pi$ and $\k^*\k$ channels are fitted separately.
The results are $1.53\pm0.23\ps$ and $2.14\pm0.40\ps$, respectively,
consistent at the level of $1.3$ standard deviations.

\subsection{Use of the tight (neural network) selection}

\begin{figure}[tbp]
  \begin{center}
  \begin{minipage}{0.9\textwidth}
     \begin{center}
       \epsfxsize=\textwidth
       \epsffile{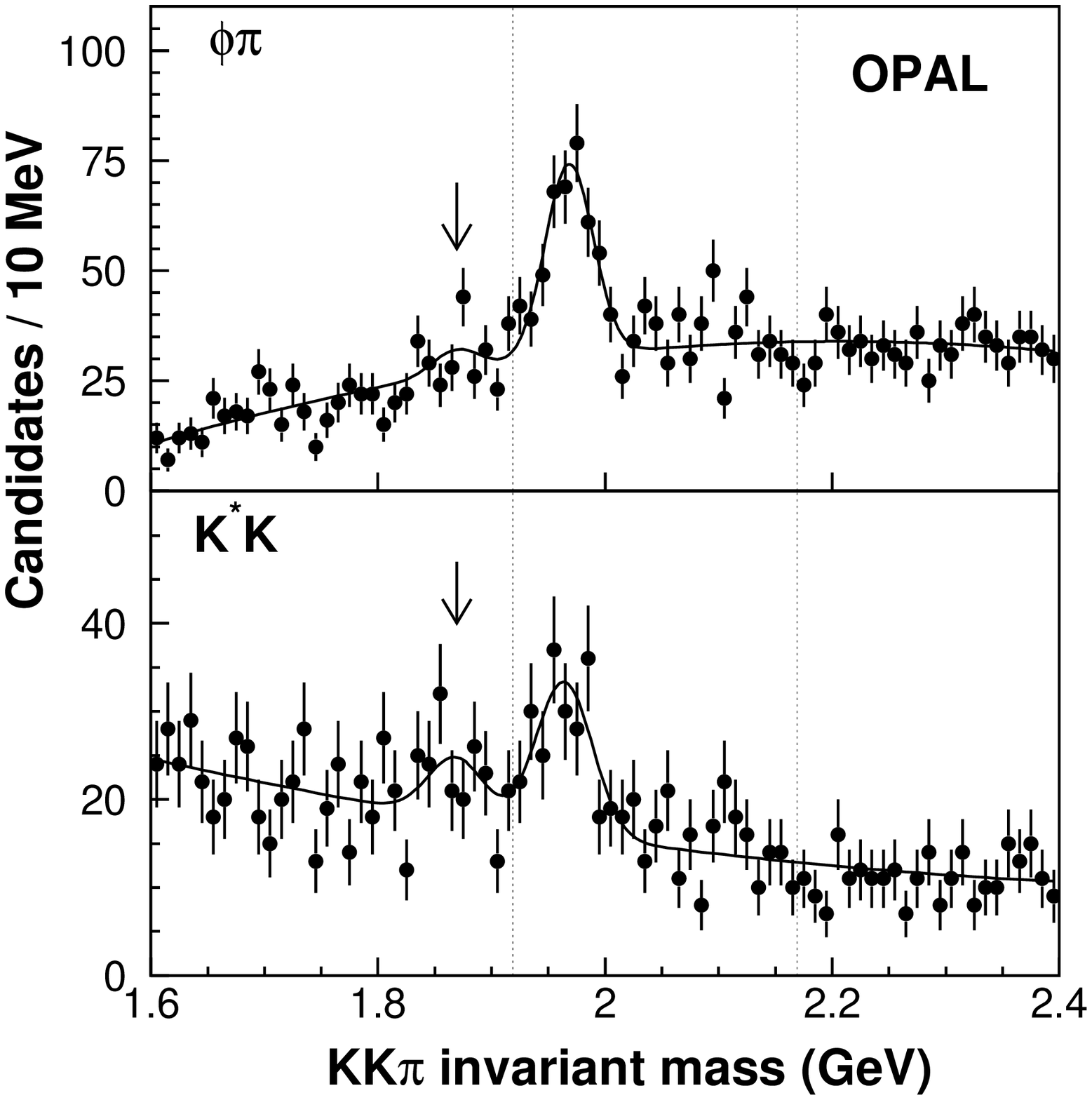}
     \end{center}
     \vspace{-0.6in}
  \end{minipage}
  \begin{minipage}{0.9\textwidth}
    \Caption{Results of the tight (neural network) selection, used as
      a cross-check of the analysis.
      Top: $\kkp$ invariant mass distribution
      for $\phi\pi$ combinations along with the fitted distribution.
      Bottom: $\kkp$ invariant mass distribution for $\k^*\k$
      candidates, along with the fitted distribution.
      The dotted lines indicate the region within which 
      candidates are used in the lifetime fit for the check.
      The arrow indicates the $\rm d^{-}$ peak position}
     \label{fig:dsmass_nn}
  \end{minipage}
  \end{center}
\end{figure}

As a check, a much tighter selection was developed which employs a
neural network to reject significantly more combinatorial background,
thereby producing a much purer $\ds$ signal.  However, this also
results in a rather significant loss of signal, and as such is not as
statistically powerful for the $\bs$ lifetime determination.  This
artificial neural network uses 16 kinematic and particle-identification
quantities, including b-tagging information in the hemisphere opposite
to the $\ds$ candidate.
The $\kkp$ invariant mass distributions for all candidates which pass
this selection are shown in figure~\ref{fig:dsmass_nn}.
The likelihood fit to the measured $\kkp$ invariant mass of the $\ds$
candidates is performed as above, resulting in a smaller signal of 
$\nnphipifitncand \pm \nnphipifitncanderr$ 
($\nnkskfitncand   \pm \nnkskfitncanderr  $) candidates
with a much reduced background fraction of 
\nnphipifitbgfrac (\nnkskfitbgfrac) for the $\phi\pi$ ($\k^*\k$) mode.


The sample of $\ds$ candidates found by the tighter neural network
selection has also been used to fit for the $\bs$ lifetime.  This sample 
gives a value of  $1.69\pm0.27\ps$ for the $\bs$ lifetime 
and is consistent with the more precise result from the analysis of
section 4. The difference in statistical
precision between the two fits  is in agreement with expectation from
toy Monte Carlo studies, given the
relative size of the $\ds$ signal and the level of combinatorial
background in each sample.

The results of all of the tests show no indication of a  
significant bias in the selection and fitting procedures.

\section{\label{sec-sys} Evaluation of Systematic Errors}
 
Systematic errors arise from  the level,
parameterization and source of the background, the potential bias from
the selection and fitting procedure, the boost estimation method, the
beam spot determination and possible tracking errors.  The systematic
errors are summarized in table~\ref{tab-syserr}.
 
\noindent{\bf Combinatorial background}

We consider the effects of both the level of the combinatorial
background, as determined by a fit to the $\kkp$ mass spectrum, and the
effective lifetime of this background, as estimated from the candidates in the
sideband region.  The systematic error due to the level of this
background is determined by repeating the calculation of the
event-by-event probability that an event is combinatorial background
by changing the estimated size of the $\ds$ signal
by the statistical uncertainty  from the fit to the
invariant mass spectrum.    
This yields a variation in the $\bs$ lifetime of
$\pm0.08\ps$.  The width of the $\kkp$ mass region from which
candidates are selected for use in the lifetime fit has also been
varied.  This was done by selecting candidates in regions extending
from 150 to 350 MeV above, and from 25 to 75 MeV below, the $\ds$
mass. Another check included using a sideband from 200 to 350~MeV
above the $\ds$ mass in place of the standard sideband from 50 to
200~MeV.  These change the fitted $\bs$ lifetime by
$\pm 0.07\ps$ which is assigned as a systematic error.  Tests
were also conducted using the toy Monte Carlo which indicated that the
level of these observed changes is consistent with the expected
$0.05\ps$ uncertainty due to statistical fluctuations in the sideband
sample; however we conservatively retain the observed variation as
a systematic uncertainty.

Several alternative parameterizations describing of the decay length
distributions of the combinatorial background have been investigated.
For example, we have included an exponential on the negative decay
length side, in place of the second Gaussian function, to describe
those events that are significantly mis-measured.  The resulting $\bs$
lifetime is $+0.02\ps$ higher than the standard result. In another
check the second wider Gaussian was used only for the combinatorial
background term in the likelihood function, changing the lifetime by
$+0.05\ps$.  Leaving out this second Gaussian altogether decreased the
lifetime by $0.10\ps$, although the quality of this fit is
significantly worse, as illustrated by an increase in $\chi^2$ of
about 300 for 40 bins, c.f. 54.7 for 42 bins with the default 
parameterization.
Consequently this last case is not considered as a systematic error.  In no
case do these alternative parameterizations significantly improve the quality
of the $\Bs$ lifetime fit, and an error of $\pm0.05\ps$ was assigned
to cover such effects.

We assign a total systematic error due to the combinatorial
background, parameterization and source, of $\pm0.12\ps$.

\noindent{\bf Physics background: sources of $\Ds$ mesons}

The physics background composition has already been discussed.   
Varying the fraction of $\Ds$ mesons from
$\Ztocc$ events over the range given in Table~\ref{tab-physbg}
produces a change of $^{+0.12}_{-0.11} \ps $ in the $\Bs$ lifetime.  
The uncertainty due to the $\Ds$ fraction from b-hadrons 
other than $\Bs$ is evaluated in the same way. The observed change 
on the $\Bs$ lifetime is $^{+0.04}_{-0.03}\ps$. 
These variations also change the statistical error 
on the fitted $\Bs$ lifetime  by up to $\pm0.03\ps$,
and this systematic effect is not considered further.

The $\Bs$ lifetime dependence on the uncertainty of the b-hadron
lifetimes has been measured by varying the b-hadron lifetimes within
the errors quoted in reference~\cite{PDG}, assuming, conservatively,
that the individual lifetimes are fully correlated with each other. The
$\Bs$ lifetime changed by $\pm0.02\ps$, which is included as a
systematic error. The $\Ds$ lifetime has also been varied in the fit,
within the errors quoted in reference~\cite{PDG}. The effect on the
measured $\Bs$ lifetime is $\pm0.01\ps$.  This variation
is also taken as a contribution to the systematic error.

The $\Ds$ momentum spectrum also depends on the fragmentation in
$\Ztocc$ and $\Ztobb$ events. Changes in the fragmentation affect the
composition of the $\Ds$ signal through changes in the efficiencies
for the charm and bottom contributions to the $\Ds$ signal (the effect
of the momentum spectrum of the $\Ds$ in $\Ztobb$ events on the
estimation of the $\Bs$ momentum in the lifetime fit is discussed
below).  In $\Ztobb$ events, we have varied the average b hadron
energy by the measurement errors~\cite{lep-avg-x_b} to yield a variation
in the observed lifetime of $\pm0.01\ps$.  Similarly, the momenta of
the $\Ds$ mesons produced in $\Ztocc$ events was varied according to 
the average
energy measured for non-strange mesons in $\Ztocc$ events (on the
assumption that there is little difference in fragmentation amongst
the various charm mesons)~\cite{opalcprod}, producing a variation of
$\pm0.01\ps$ in $\tauBs$.

In the case of $\Ds$ mesons coming from the decay of b-hadrons other
than $\Bs$, the $\Ds$ may be produced in either a two-body mode (e.g.
$\B\to\Ds\D$) or a multi-body final state where one or more light
particles are produced.  The two-body decay fraction of $\B\to\Ds\X$
has been measured to be $0.56\pm0.10$~\cite{BtoDs-2body}.  In
determining the relative efficiency of these $\Ds$ mesons from
b-hadrons other than $\Bs$, we have already corrected our simulation
to this two-body fraction.  Assuming b-baryons decaying to $\Ds$
mesons in two- or multi-body states have the same fractions as $\Bu$
and $\Bd$ mesons, the efficiency of selecting $\Ds$ mesons from these
decays has been evaluated by varying the two-body fraction over the
range 0.46 to 0.66 and re-evaluating the relative efficiency of this
source of $\Ds$ with respect to the $\Ds$ which arise from a $\Bs$
decay.  This produces a variation of $\pm0.01\ps$ in the fitted $\Bs$
lifetime.

Thus we assign a total error due to these other sources of $\Ds$
mesons of $^{+0.13}_{-0.12}\ps$.

\noindent {\bf Boost estimation}

The energy spectrum of the Monte Carlo $\Bs$ events used to estimate
the momentum of the $\Bs$, given the observed $\Ds$ momentum, can also
affect the resulting lifetime.  This effect is not large because the
scaling used to estimate the $\Bs$ momentum from the measured $\Ds$
momentum is correlated with the $\Ds$ momentum itself.  We have varied
the average $\Bs$ energy by the measured errors on the average
b-hadron energy~\cite{lep-avg-x_b} to yield a variation in the $\Bs$
lifetime of $\pm 0.01\ps$.\footnote{Note that the uncertainty due to
  the bottom and charm hadron energy spectra affects the $\Bs$
  lifetime both through the boost estimation and through the $\Ds$
  sample composition.  When combined, these two contributions are
  added linearly.} The effect of a $2.0\,\MeV$ uncertainty in the
mass of the $\Bs$~\cite{PDG} was found to produce a change of less
than $0.01\ps$ in the $\Bs$ lifetime.

\noindent {\bf Beam position and size} 

The average intersection point of the LEP beams in OPAL is used to
estimate the production vertex of the $\Bs$ candidates.  The
sensitivity of \tauBs\ to the assumed position and size of the beam
spot has been evaluated as in reference~\cite{opalbslife}, resulting in
a variation in the fitted lifetime of no more than $0.01\ps$, which has
been assigned as a systematic error.

\noindent {\bf Detector alignment} 

The effects of alignment and calibration uncertainties on the $\Bs$
lifetime are estimated from a detailed study of 3-prong
$\mathrm{\tau}$ decays~\cite{taulife}.  These uncertainties lead to an
uncertainty of $0.01\ps$ on $\tauBs$.
\begin{table}[thb]
  \centering
  \begin{tabular}{|l|r|r|}
    \hline
      source  & uncertainty (ps)\\
    \hline
      combinatorial background &  $\pm 0.12$ \\
      physics background       &  $^{+0.13}_{-0.12}$ \\
      uncertainty in boost     &  $\pm 0.01$\\
      beam spot                &  $\pm 0.01$\\
      alignment errors         &  $\pm 0.01$\\
    \hline
      total                    &  $\taubsys$\\
    \hline
  \end{tabular}
  \Caption{Summary of systematic errors on the $\Bs$ lifetime.}
  \label{tab-syserr}
\end{table}
 
Combining the systematic errors in
Table~\ref{tab-syserr}, 
we find
$\tauBs = \fittaubs 
(\mathrm{stat})\taubsys(\mathrm{syst})\ps.$

\section{Conclusion}
 
A sample of  $\Dsm$ decays has been reconstructed in
which the $\Dsm$ has decayed into $\KKpi$ through either the
$\phi\pi^-$ or $\K^{*0}\K^-$ channels.  From \nGPMH~million hadronic
$\Zzero$ events recorded by OPAL from 1991 to 1995, a total of $\ncand
\pm \ncanderr$ such candidate decays have been found, of which about
57\% are expected to be from $\Bs$ decay.  The $\Bs$ lifetime is found
to be
\[ \tauBs   = \fittaubs (\mathrm{stat}) \taubsys
(\mathrm{syst}) \ps, \]   
a result consistent with the measured $\rm B^0$ lifetime and other
$\Bs$ measurements \cite{PDG}. This result is also
in agreement with current theoretical expectations \cite{BIGI,NEUBERT}.
 
\par
Acknowledgements:
\par
We particularly wish to thank the SL Division for the efficient operation
of the LEP accelerator at all energies
 and for
their continuing close cooperation with
our experimental group.  We thank our colleagues from CEA, DAPNIA/SPP,
CE-Saclay for their efforts over the years on the time-of-flight and trigger
systems which we continue to use.  In addition to the support staff at our own
institutions we are pleased to acknowledge the  \\
Department of Energy, USA, \\
National Science Foundation, USA, \\
Particle Physics and Astronomy Research Council, UK, \\
Natural Sciences and Engineering Research Council, Canada, \\
Israel Science Foundation, administered by the Israel
Academy of Science and Humanities, \\
Minerva Gesellschaft, \\
Benoziyo Center for High Energy Physics,\\
Japanese Ministry of Education, Science and Culture (the
Monbusho) and a grant under the Monbusho International
Science Research Program,\\
German Israeli Bi-national Science Foundation (GIF), \\
Bundesministerium f\"ur Bildung, Wissenschaft,
Forschung und Technologie, Germany, \\
National Research Council of Canada, \\
Hungarian Foundation for Scientific Research, OTKA T-016660, 
T023793 and OTKA F-023259.\\

%
 

\end{document}